# Structure, microstructure and electrical properties of new lead-free $(1-x)(Li_{0.12}Na_{0.88})NbO_3$-$xBaTiO_3$ ($0 \leq x \leq 40$) piezoelectric ceramics

Supratim Mitra[*], Ajit R. Kulkarni[#]

*Department of Metallurgical Engineering and Materials Science, Indian Institute of Technology Bombay, Mumbai 400076, India*

[*]*Now at: Department of Basic Sciences, NIIT University, Neemrana, Rajasthan 301705, India*

[#]Corresponding Author: Ajit R. Kulkarni
Tel.: +91 22 2576 7636;  Fax.:+91 22 2572 3480
E-mail address: ajit.kulkarni@iitb.ac.in

## Abstract

A new lead-free $(1-x)Li_{0.12}Na_{0.88}NbO_3$-$xBaTiO_3$ ($0 \leq x \leq 40$) piezoelectric ceramics have been prepared using conventional ceramics processing route. Structural analysis revealed an existence of morphotropic phase boundary (MPB), separating orthorhombic and tetragonal phases, between the $BaTiO_3$ content, $x$ = 10-12.5. A partial phase diagram has been established based on temperature-dependent permittivity data for this new system and a nearly vertical temperature-independent MPB is observed. An improvement in electrical properties near MPB (e.g., for $x$ = 12.5; $\varepsilon_r$ = 10489 at $T_m$, $d_{33}$ = 30 pC/N, $k_p$ = 12.0 %, $Q_m$ = 162, $P_r$ = 11.2 μC/cm$^2$, $E_c$ = 19.2 kV/cm, $d^*_{33}$ = 269 pm/V) is observed and is attributed to the ease of polarization rotation due to coexistence of orthorhombic and tetragonal phases. The results show that these materials could be suitable for piezoelectric vibrators and ultrasonic transducers applications. The sample with $x$ = 25, also shows high dielectric permittivity, $\varepsilon_r$ = 3060, and low dielectric loss, $tan\delta$ = 0.033 which could be suitable for capacitor (X7R/Z5U) applications.

**Keywords:** Piezoelectric Ceramics; Morphotropic Phase Boundary; Dielectric Permittivity; Microstructure, Phase Diagram.

**1. Introduction**

Pb(Zr,Ti)O$_3$ (PZT)-based piezoelectric ceramics are widely used in electronic equipment such as sensors, actuators, transducers, capacitors due to their excellent electrical properties[1, 2]. High piezoelectric and ferroelectric properties in these systems have long been recognized to arise from the coexistence of thermodynamically equivalent phases at the morphotropic phase boundary (MPB) composition. It has been widely reported that in PZT, MPB is a nearly vertical phase boundary that separates rhombohedral, tetragonal, and/or monoclinic phases where a high degree of alignment of ferroelectric dipoles takes place and results in dramatic enhancement in the ferroelectric properties. However, recently researchers have been focusing on lead-free piezoelectric materials because the toxicity of lead (Pb) and its compound, that are used in the most of these electronic devices. Therefore, an urgent need to develop lead-free piezoelectric ceramics as an alternative of PZT is realized. Nevertheless, it is true that there is no easy option for PZT and the research in the field of new lead-free piezoelectric ceramics should emphasis on specific device applications.

In order to explore new lead-free piezoelectric materials, the most extensively studied lead-free piezoelectric materials with perovskite structure have been focused on (Bi$_{0.5}$Na$_{0.5}$)TiO$_3$ (BNT)- and (K$_{0.5}$Na$_{0.5}$)NbO$_3$ (KNN)-based systems and their MPB compositions[3, 4]. BNT is one of the important lead-free piezoelectrics because of strong ferroelectric properties (P$_r$ = 38 μC/cm$^2$ ). However, pure BNT have some major drawbacks e.g., low depolarization temperature (T$_d$ < 150 °C), high conductivity, higher coercive field (E$_C$ = 73 kV/cm), and higher sintering temperature (> 1200 °C)[5]. KNN has two serious problems, firstly low density for the sintered ceramics, by conventional cost effective route and secondly low mechanical quality factor ($Q_m$)[6-9]. As a result, the usage of these materials is limited to many practical applications. Hence, in order to resolve this, many additives are incorporated with these ceramics systems. For example, BNT is altered with $(Bi_{0.5}K_{0.5})TiO_3$, $BaTiO_3$, $Ba(Zr,Ti)O_3$ etc. to form binary and ternary systems, while KNN is modified with $LiSbO_3$, $BaTiO_3$ etc [5, 10]. Although new BNT-based systems showed good piezoelectric properties near its MPB compositions, their depolarization temperature is still low ($T_d \sim 150$ °C). KNN-based systems show high piezoelectric properties by forming polymorphic phase transition (PPT) near room-temperature. However, the electrical properties of all these systems are comparatively poor and unable to replace PZT completely. Therefore, there is a demand for searching new high performance lead-free piezoelectric ceramics preferably with a MPB composition, as existence of MPB are accepted to enhance dielectric, piezoelectric, and ferroelectric properties.

Another recently studied important lead-free piezoelectric system with perovskite structure, $(Li_{0.12}Na_{0.88})NbO_3$ with an MPB (LNN), is known for resonators/filters applications due to its high ($Q_m$) and low dielectric loss. However, LNN also have drawbacks e.g., higher sintering temperature, low piezoelectric properties/constant ($d_{33}$), and poor temperature stability. Therefore, in order to improve piezoelectric properties of LNN-based ceramics, various modifications were made with K, Sb, Ta by many researchers[11-14] and a significant improvement in $Q_m$ has been found. However, problems with high sintering temperature, low piezoelectric properties and temperature stability remain unimproved. Also, the effect of addition of another perovskite component to LNN is hardly reported. On the other hand, $BaTiO_3$ (BT), a typical perovskite ferroelectric system, has been recognized for its high dielectric, piezoelectric properties and low dielectric loss but low $T_C$ limits its practical applications. Interestingly, BT with tetragonal structure at room-temperature is a feasible end member that can form an MPB with other niobate-based end members e.g., $KNbO_3$, $NaNbO_3$,

AgNbO$_3$, and (K,Na)NbO$_3$ ceramics with orthorhombic structure which also enhances the piezoelectric properties of these ceramics [15-19].

In view of the above discussion, a new solid-solution between niobate-based LNN and BT, (1-$x$)LNN-$x$BT were selected to achieve high-performance piezoelectric ceramic system. In this work, (1-$x$)LNN-$x$BT (0 ≤ $x$ ≤ 40) were prepared using conventional ceramic route with an aim to improve the piezoelectric property of LNN, enhance densification along with a reduction in the sintering temperature, and formation of an MPB composition. The phase structure, microstructure, and electrical properties were investigated in details during this work.

## 2. Experimental

Solid solution of (1-$x$)Li$_{0.12}$Na$_{0.88}$NbO$_3$-$x$BaTiO$_3$ [(1-$x$)LNN-$x$BT] ($x$ = 0, 5, 7.5, 10, 12.5, 15, 20, 25, 30, 40 mol%) were prepared using conventional solid state reaction followed by sintering. Here, the BT content ($x$ mol%) was restricted to 40% as when $x$ ≥ 40, the Curie-temperature became too low (< -100 $^o$C) for any practical use. The starting raw materials for the synthesis were reagent-grade TiO$_2$, Nb$_2$O$_5$, Na$_2$CO$_3$ (both 99.5% pure, Loba Chemie, India), Ba$_2$CO$_3$ (99.0% pure, Aldrich, USA) and Li$_2$CO$_3$ (99.0% pure, Merck, India). LNN and BT powders were synthesized separately from the raw materials Nb$_2$O$_5$, Na$_2$CO$_3$, Li$_2$CO$_3$ and TiO$_2$, Ba$_2$CO$_3$ respectively (details are given elsewhere [2, 20]). These powders were mixed in the desired stoichiometry of (1-$x$)LNN-BT (0 ≤ $x$ ≤ 40) and ball milled for 24h followed by compaction into pellet and pressureless sintering at different temperatures between 1100-1250 $^o$C for 3h. The phase purity of the ceramic samples was checked by powder X-Ray Diffraction (XRD) at room temperature using X-ray diffractometer (X'Pert, PANalytical) with Cu-K$\alpha$ radiation. For this, the final sintered pellets were finely crushed and annealed at 500 $^o$C for 12h to reduce residual strain introduced by crushing. For microstructural analysis, the sintered pellets were polished and thermally etched. The microstructure was recorded using scanning electron microscope (SEM) (JEOL-JSM 7600F). The average grain size, $G$, was estimated using lineal intercept data collected over 200-300 grains, sampled at different locations of each of the SEM micrographs, and following the image analysis software (Image J, NIH, USA). For electrical measurements, silver paste was fired on both sides of the pellet samples at 500 $^o$C for 30 min. The dielectric constant/permittivity of the unpoled samples was recorded using Impedance Analyzer (Alpha High Resolution, Novocontrol, Germany) in the frequency range 0.01 Hz-10 MHz over the temperatures -100 to 500 $^o$C. The planar electromechanical coupling factor $k_p$ and mechanical quality factor $Q_m$ were determined on the poled samples using impedance analyzer via resonance-antiresonance method. For this, the disk samples were poled at different temperatures (50-150 $^o$C) under the DC field of 4-7 kV/mm for 30 min. The piezoelectric constant $d_{33}$ was measured using $d_{33}$-meter (Piezotest, PM300, UK).

## 3. Results and discussion

### 3.1. XRD analysis

Figure 1(a) shows the X-ray diffraction (XRD) pattern of (1-$x$)LNN-$x$BT ceramics at room temperature as a function of BT content ($x$). It can be seen from the figure 1(a) that all samples show a pure perovskite phase, which confirms the formation of homogeneous solid-solutions for the studied composition range $0 \leq x \leq 40$. This can also be explained as follows: the radii of $Ba^{2+}$ (1.47 Å, CN = 9) is close to $Na^+$ (1.24 Å, CN = 9) and $Li^+$ (~1.00 Å, CN = 9) and much larger than $Nb^{5+}$ (0.64 Å, CN = 6). Therefore, $Ba^{2+}$ is suitable for $A$-site substitution of the $ABO_3$ perovskite structure. Conversely, radius of $Ti^{4+}$ (0.605 Å, CN = 6) is very close to the radius of $Nb^{5+}$ and is introduced in $B$-site. However, a minor secondary $Ba_6Ti_2Nb_8O_{30}$ phase has also been observed.

To analyze the structure, all the peaks were indexed in terms of cubic perovskite unit cell. The selected pseudocubic (110) and (200) peaks shown in figure 1(b) and 1(c) are found to shift systematically towards lower 2θ position with increase in BT content ($x$) for compositions $10 \leq x \leq 40$ confirming a homogeneous solid solution. This also indicates that larger $Ba^{2+}$ substitutes smaller $Na^+$/ $Li^+$ and increases the cell volume as can be seen in table 1. The lattice parameters and the unit cell volumes for all the compositions ($0 \leq x \leq 40$) are calculated and listed in table 1.

It can also be seen from Figure 1(b) and 1(c), that samples with compositions $x > 20$ show cubic (C) structure with space group $Pm\bar{3}m$ characterized by all singlet reflections. For the compositions $x \leq 20$, the structure become tetragonal (T) characterized by doublet (110) and (200) and singlet (111) with space group *P4mm*. The tetragonal structure remains up to $x = 12.5$ and then changes to an orthorhombic (O) structure of LNN phase with space group *Pbma* consistent with our earlier results [20]. Thus two phase transitions exist in (1-$x$)LNN-$x$BT for the studied composition ranges $0 \leq x \leq 40$. A transformation from C-phase to T-phase is identified near $x = 20$. An abrupt structural change with compositions defined as morphotropic phase boundary (MPB) separating orthorhombic (O) and tetragonal (T) ferroelectric phase is believed to exist between $x = 10$ to 12.5. In fact, the so-called MPB in KNN-based systems is a ferroelectric orthorhombic-tetragonal PPT and shows strong temperature dependence in temperature-composition phase diagram[21, 22].

## 3.2. Densification and microstructural evolution

In structural analysis, the secondary phase ($Ba_6Ti_2Nb_8O_{30}$) appeared as a common observation of alkali niobate-based ceramics sintered at elevated temperature [23]. These secondary phases could be related to volatilization of $Li_2O$ and $Na_2O$. The effect of this volatilization can be seen in microstructure evolution and densification process. The relative density (%) and optimum sintering temperature for (1-$x$)LNN-$x$BT (0 ≤ $x$ ≤ 40) ceramics are shown in figure 2(a). For each composition ($x$), the ceramics were sintered at different temperatures and simultaneously the sintered density was measured using standard method (ASTM). The optimum sintering temperatures for a composition is decided as the sintering temperature by which the sample shows the highest relative density [24]. As shown in figure 2(a), the densification is found to improve with addition of BaTiO$_3$ ($x$), while sintering temperature is reduced in (1-$x$)LNN-$x$BT ceramics and can be understood by the following. The maximum relative density (%) and the corresponding optimum sintering temperature are listed in table 2. It can be seen from table 2 and figure 2(a) that due to addition of BaTiO$_3$ ($x$), the relative density has increased from 95.30% for $x$ = 0 to 97.55% for $x$ = 10 and then decreased very slightly with further increase in $x$. However, for all the BaTiO$_3$ added samples, relative density were found in the range of ~96.5-97.5%, while the optimum sintering temperature decreased from 1250 $^o$C (for $x$ = 0) to the ranges of 1185-1120 $^o$C as seen in table 2 and figure 2(a).

Figure 3 shows the SEM micrographs of polished and thermally etched surfaces of (1-$x$)LNN-$x$BT ceramics for $x$ = 5 to 40 where, it can also be seen that BaTiO$_3$ content ($x$) has a significant role in microstructure evolution. For all the samples, dense and well developed grains are seen, albeit, with a non-uniform grain size distribution. A careful observation of the grain size distribution histograms, given as insets in figures 3, reveals bimodal distributions. The characteristics of bimodal grain-size distribution, consisting of a large fraction of smaller grains and a small fraction of larger grains, increased as $x$ increased from 5 to 12.5 and then gradually became uniform as $x$ increased from 12.5 to 40.

In (1-$x$)LNN-$x$BT ceramics, liquid phase assisted densification and grain growth is expected as BaTiO$_3$ content increases leading to a non-uniform grain size distributions. The exact reason for this grain growth is still unknown however, it is believed that it is a consequence of volatilization of $Li_2O$ and $Na_2O$ which leads to formation of Li/Na deficient $Ba_6Ti_2Nb_8O_{30}$, a low melting secondary phase at the grain boundary as evidenced from XRD [23, 25-29]. As BT content increases from $x = 5$ to 12.5, the secondary phase start appearing which leads to liquid phase sintering and promote grain growth and consequently the average grain size, $G$ has increased from 2.1 to 6.5 μm. Further addition of $BaTiO_3$ ($x$) limits the grain growth by restricting the mobility of grain boundary causing inhibited grain growth, and consequently the average grain size has gradually decreased to 2.1 μm as $x$ reaches to 40 as shown in figure 2(b)[29].

### 3.3. Temperature-dependent dielectric permittivity

The temperature-dependence (50 to 500 °C) of the dielectric constant and dielectric loss measured at 1 kHz for (1-$x$)LNN-$x$BT (0 ≤ $x$ ≤ 40) ceramics is shown in figure 4(a) and (b) respectively. The inset in figure 4(a) and (b) shows the same plot for temperature range of -100 to 100 °C for the composition $x$ = 25, 30, 40. Samples with $x$ = 0 to 15 show two dielectric peaks, associated with the phase transition of orthorhombic-tetragonal ($T_{O-T}$) and tetragonal-cubic ($T_m$), respectively. Both the transition temperatures ($T_{O-T}$ and $T_m$) are found to decrease with increase in BT content ($x$) while an orthorhombic-tetragonal phase transition ($T_{O-T}$) has not been observed for $x$ = 20 to 40, as they shifts below -100 °C. The decrease in transition temperatures towards lower temperature side as $x$ ($BaTiO_3$ content) increases is because of the valence mismatch of both $A$ and $B$-site cation [9]. Usually, the $T_{O-T}$ located near or below room temperature shows good temperature stability of alkali niobate based ceramics [15, 18, 30].

The dielectric constant maxima at transition temperature $\varepsilon_r(T_m)$ is found to increase from 2703 for $x = 0$ up to 10489 for the MPB composition $x = 10$ and then decreases on further increase in $x$ as can be seen in table 3. The substitution of large $Ba^{2+}$ ion for small $Na^+/Li^+$ ion increases the rattling space of oxygen-octahedra, increasing the polarizability of electric dipoles and dielectric constant. The increased grain size near MPB composition is also responsible for high dielectric constant. Interestingly, it has been observed that width of the dielectric peaks varies with $x$ and increases with increase in $x$, which has been discussed in details in another report [31]. Larger the width of the dielectric peak, larger is the temperature stability. Therefore, a broad dielectric peak with a higher dielectric constant ($\varepsilon_r \sim 3000$) along with low dielectric loss as listed in table 3 indicate that (1-$x$)LNN-$x$BT ceramics would be a good candidate for various capacitor applications.

### 3.4. Phase diagram

Determining the transition temperatures ($T_m$, $T_{O-T}$) from the temperature-dependent dielectric permittivity as shown in figure 4, a partial phase diagram has been proposed for (1-$x$)LNN-$x$BT ($0 \leq x \leq 40$) ceramics and shown in figure 5. From the composition-temperature phase diagram it can be seen that the orthorhombic-tetragonal phase boundary ($T_{O-T}$) is almost temperature independent indicating an existence of a nearly vertical MPB around $x = 10$. A vertical MPB is always expected so that the temperature variation could not able to keep the materials away from MPB [32]. In case of a temperature-dependent MPB, a slight variation of temperature brought away the materials from MPB, i.e., from the instability condition and results in a decrease in properties. In PZT, nearly vertical MPB separating rhombohedral and tetragonal phases are observed [2] however, in our recent study for LNN system (without BT addition) [20] a curved MPB has been observed. In KNN-based system instead a vertical composition-temperature phase boundary, known as PPT, a strongly temperature dependent phase boundary has been observed resulting in decreased electrical properties[21, 22].

### 3.5. Ferroelectric and piezoelectric properties

Figure 6(a) shows the room temperature *P-E* hysteresis loops measured on unpoled (1-*x*)LNN-*x*BT ceramics under an applied electric field of 50 kV/cm at 1 Hz for $0 \leq x \leq 20$, while for $25 \leq x \leq 40$, the *P-E* loop measured under an applied electric field of 25 kV/cm at 1 Hz as shown in inset of figure 6(a) for a comparison. The samples with composition up to $x = 20$ show a normal hysteresis loop confirming ferroelectric nature of (1-*x*)LNN-*x*BT ceramics. For $x = 25$, a typical ferroelectric hysteresis with low remnant polarization ($P_r$) can still be observed at room temperature (which is very close to its Curie-temperature, 56 $^o$C as seen in inset of figure 4), while for $x > 25$ samples are paraelectric in nature and hence the loop almost disappears as evidenced from XRD. The variation of remnant polarization ($P_r$) and coercive field ($E_C$) with BT content (*x*) is also shown in the inset of figure 6(a). It can be seen that BT content drastically increase the $P_r$-value from 1.1 $\mu$C/cm$^2$ for $x = 0$ (LNN) to 11.2 $\mu$C/cm$^2$ for $x = 12.5$, which is near MPB composition for (1-*x*)LNN-*x*BT ceramics. The $E_C$-values also found to decrease as *x* approaches to MPB composition and found a minimum value of 17.6 kV/cm for $x = 15$. In pure BNT, a very high value of $E_C = 73$ kV/cm, while for a MPB composition of BNT-BT, $E_C$-value of 27 kV/cm is reported, which are still high as compared to found in the present study. In case of KNN-based systems, $E_C < 20$ kV/cm has been observed, however $P_r$-values are found to have a decreased value [15].

Figure 7 shows typical ferroelectric butterfly shaped *S-E* curve measured under the applied electric field of 50 kV/cm at 1 Hz for the composition range $x = 5$-20 (FE-phase at room temperature). The observed asymmetric butterfly loop in the *S-E* curve could be attributed to the internal bias field arising from defects [31, 33]. The unipolar *S-E* curves and the variation of large signal $d^*_{33}$ (defined as $S_{max}/E_{max}$) with compositions near MPB are shown in the insets of the figure 7. The maximum $d^*_{33}$ –value of 269 pm/V are found for $x = 12.5$ which are comparable to KNN and BNT-based system where $d^*_{33}$ –value ranges from 75 pm/V for $SrTiO_3$ modified KNN to 400 pm/V for Li, Ta, Sb modified KNN and 240 pm/V for BNT-BKT binary system to 567 pm/V for BNT-BT-BKT ternary system [30]. The variation of remnant polarization $P_r$, coercive field $E_c$ and large-signal $d^*_{33}$ of the (1-*x*)LNN-*x*BT ceramics at different *x* are summarized in table 3.

The variation of piezoelectric coefficient ($d_{33}$), planar coupling constant ($k_p$), and mechanical quality factor ($Q_m$) of (1-*x*)LNN-*x*BT ceramics as a function of BT content (*x*) are listed in table 3. A nominal increase in the $d_{33}$-values are found as *x* approaches to MPB, however a maximum value of 49 pC/N is seen for $x = 20$, whereas, a decreasing trend in $k_p$ and $Q_m$ are found as *x* increases from $x = 0$ to 40. The range of minimum to maximum values of $Q_m$ that obtained in (1-*x*)LNN-*x*BT ceramics near MPB ($x = 7.5$-15) is 129 to 267, which is still superior than BNT and KNN-based system [10].

## 4. Summary

In summary, a new lead-free system, (1-*x*)LNN-*x*BT have been selected and prepared using conventional ceramic route. The phase and microstructure analysis confirm a homogeneous solid solution and an existence of MPB composition near $x = 10$-12.5. Temperature-dependent dielectric permittivity studies show a nearly temperature-independent vertical MPB in composition-temperature phase diagram. Improved electrical properties near MPB (e.g., for $x = 12.5$; $\varepsilon_r = 10489$ at $T_m$, $d_{33} = 30$ pC/N, $k_p = 12.0$ %, $Q_m = 162$, $P_r = 11.2$ $\mu C/cm^2$, $E_c = 19.2$ kV/cm, $d^*_{33} = 269$ pm/V) is attributed to ease of polarization rotation due to coexistence of orthorhombic and tetragonal phases. The results show that these materials could be suitable for piezoelectric vibrators and low power transducers applications. In addition, sample with $x = 25$, shows high dielectric permittivity, $\varepsilon_r = 3060$, low dielectric loss, $tan\delta = 0.033$, a homogeneous grain size distribution with an optimum average grain size ~ 2 µm, and a diffused phase transition near ferroelectric-paraelectric transition. Therefore, (1-$x$)LNN-$x$BT ceramics with composition near $x = 25$ could be suitable for monolithic or MLCC capacitor (X7R/Z5U) applications. The results shows that (1-$x$)LNN-$x$BT ceramics could be a potential lead-free piezoelectric system for various electronic system and come with a great promises.

**Acknowledgement:**

Authors would like to thank P. K. Ojha for $d_{33}$ measurements.

**Figure caption:**

**Figure 1:** (a) XRD pattern of (1-$x$)LNN-$x$BT ceramics for $x$ = 0-40; amplified XRD pattern in the 2θ range (b) 31.5°-33.0°, (c) 45.5°-47.5° shown below. Gradual changes of phase from orthorhombic (O-Phase) to tetragonal (T-Phase) and tetragonal (T-Phase) to cubic (C-Phase) found at $x$ = 10 and 25 respectively.

**Figure 2:** Variation of (a) sintering temperature, and relative density (%) (b) Average grain size, with $x$ for (1-$x$)LNN-$x$BT ceramics.

**Figure 3:** SEM micrographs of (1-$x$)LNN-$x$BT ceramics for (a) $x$ = 5, (b) $x$ = 7.5, (c) $x$ = 10, (d) $x$ = 12.5, (e) $x$ = 15, (f) $x$ = 20, (g) $x$ = 25 (h) $x$ = 30, (i) $x$ = 40; inset shows the histogram of grain size distribution.

**Figure 4:** Plot of dielectric constant as a function of temperature for the range (a) 50 to 500 °C; (c) -100 to 100 °C, and dielectric loss as a function of temperature for the range (b) 50 to 500 °C; (d) -100 to 100 °C for (1-$x$)LNN-$x$BT (0 ≤ $x$ ≤ 40) ceramics at 1 kHz.

**Figure 5:** Temperature-composition phase diagram of (1-$x$)LNN-$x$BT (0 ≤ $x$ ≤ 40) ceramics, based on dielectric permittivity data.

**Figure 6:** P-E hysteresis loop for (1-$x$)LNN-$x$BT ceramics with different $x$. (a) for 0 ≤ $x$ ≤ 40, (b) for 25 ≤ $x$ ≤ 40 showing a change from ferroelectric to paraelectric phase . (c) Variation of remnant polarization $P_r$ and coercive field $E_c$ with $x$ in (1-$x$)LNN-$x$BT ceramics for 0 ≤ $x$ ≤ 20.

**Figure 7:** (a) Field-induced strain (S-E) curve, (b) Unipolar S-E curve (top inset), for (1-$x$)LNN-$x$BT ceramics with different $x$ (5 ≤ $x$ ≤ 20). The variation of large signal $d^*_{33}$ (defined as $S_{max}/E_{max}$) with compositions near MPB (7.5 ≤ $x$ ≤ 20) for (1-$x$)LNN-$x$BT ceramics shown in bottom inset.

**Table caption:**

**Table 1:** Variation of lattice parameters at room-temperature for (1-$x$)LNN-$x$BT ceramics with composition $x$.

**Table 2**: Variation of maximum relative density (%) measured at corresponding optimum sintering temperature ($^o$C) at different $x$ in (1-$x$)LNN-$x$BT ceramics.

**Table 3:** The electrical properties of (1-$x$)LNN-$x$BT ceramics ($0 \leq x \leq 40$).

**Figure 1**

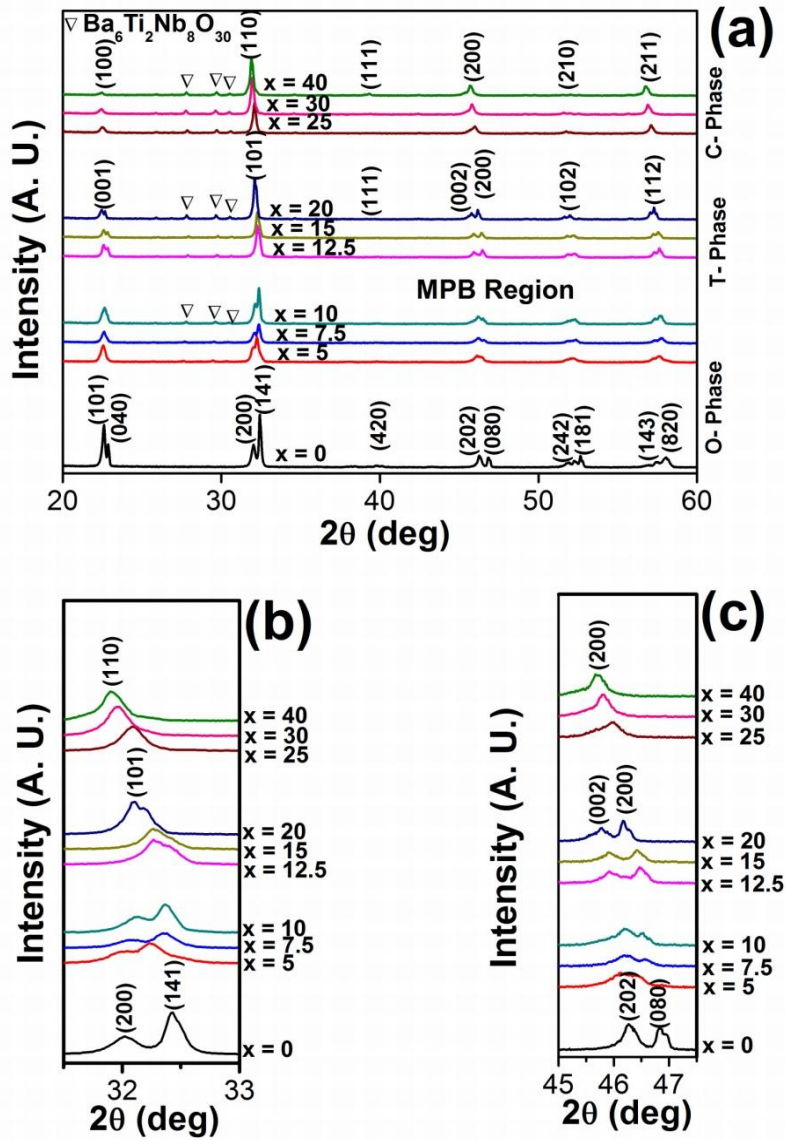

**Figure 2**

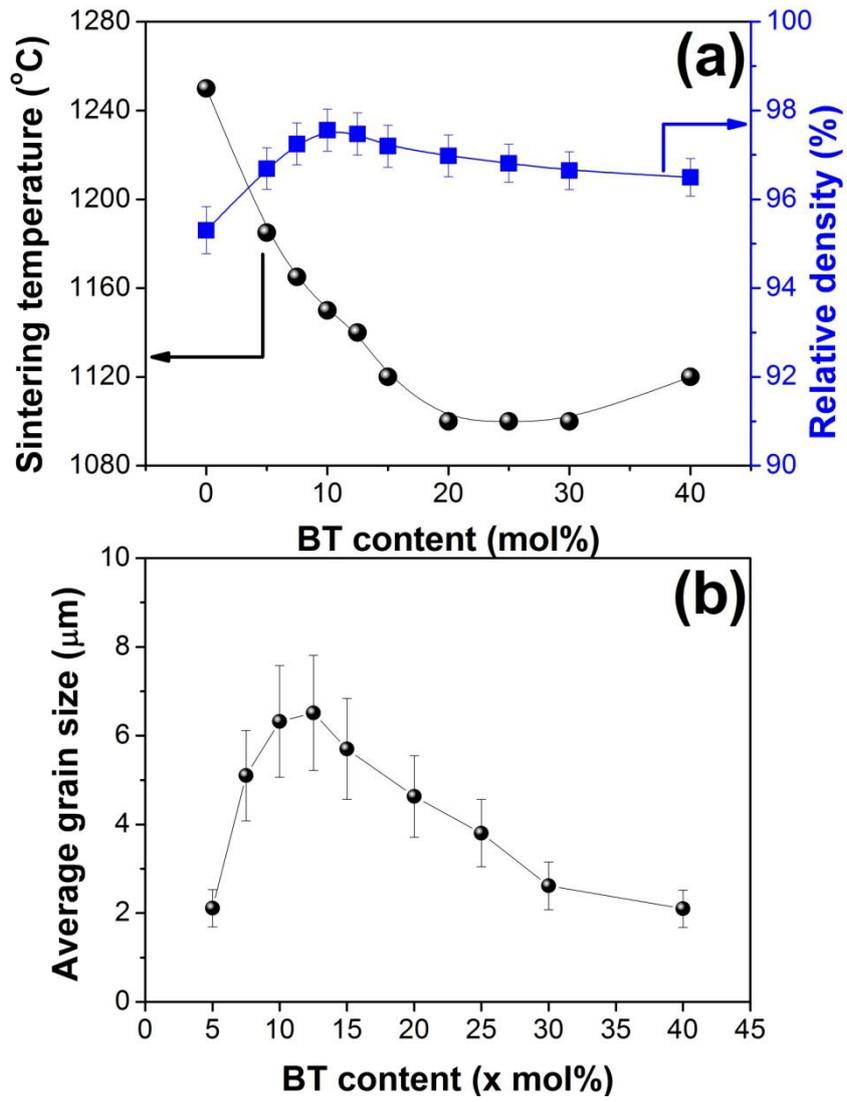

**Figure 3**

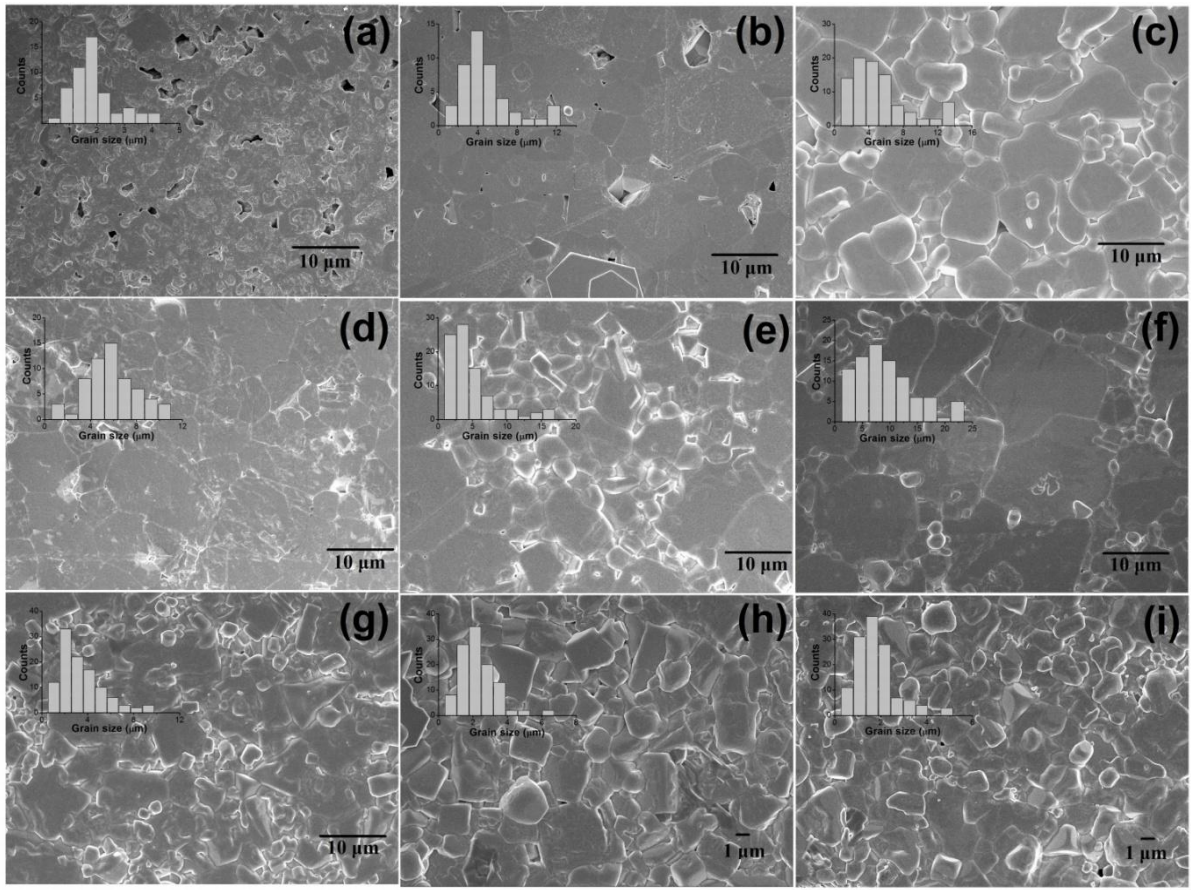

**Figure 4**

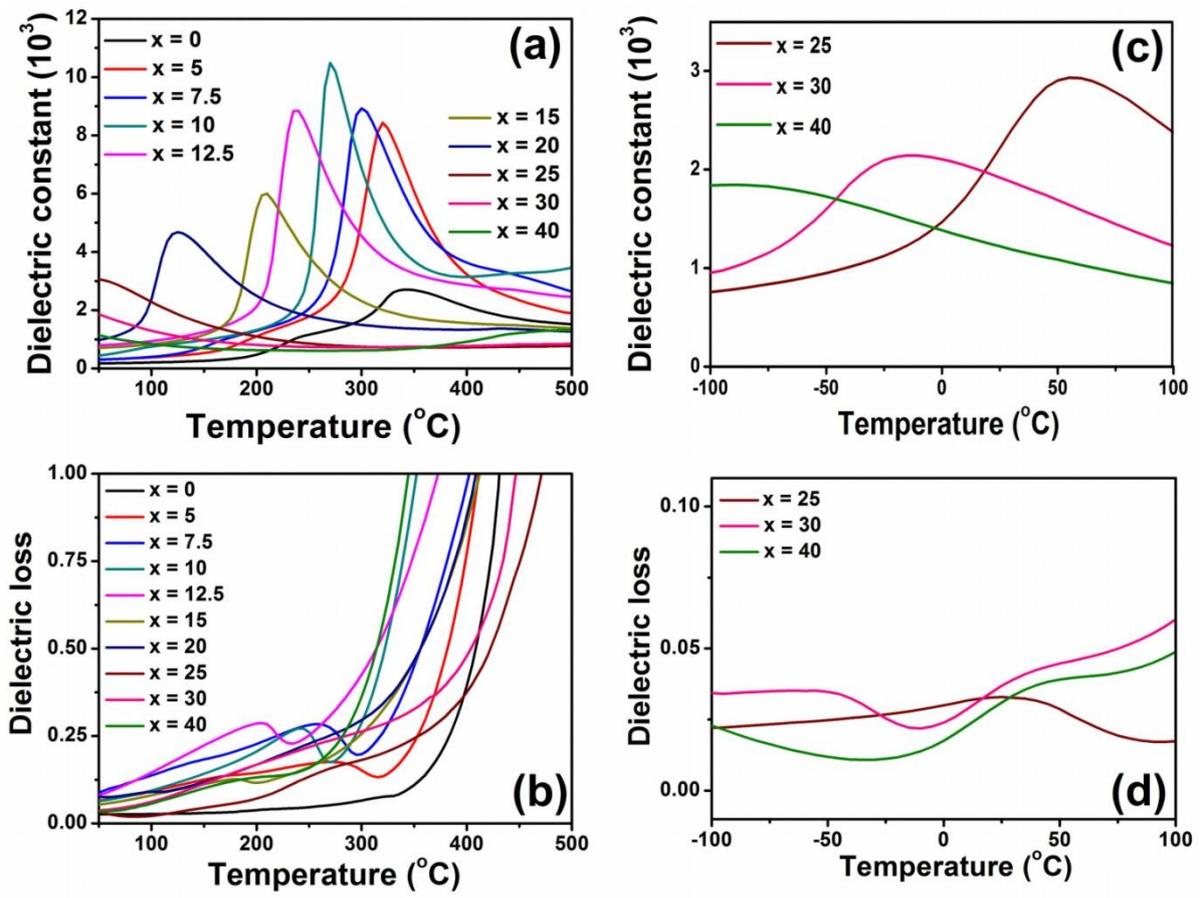

**Figure 5**

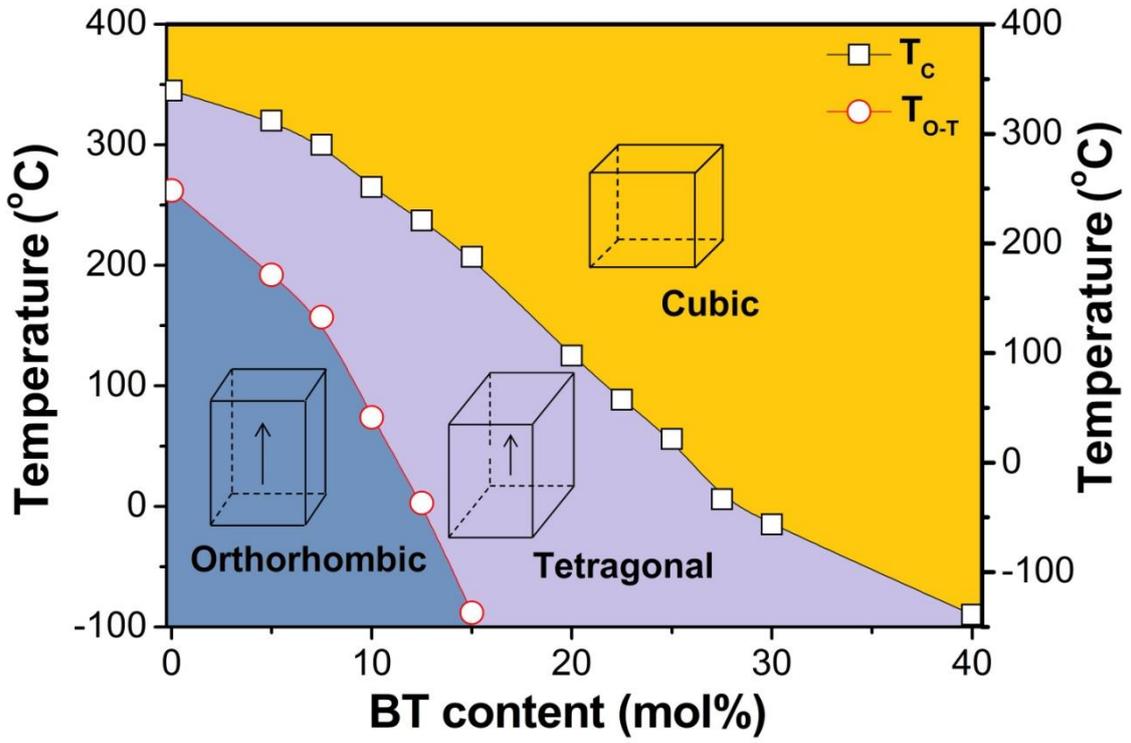

**Figure 6**

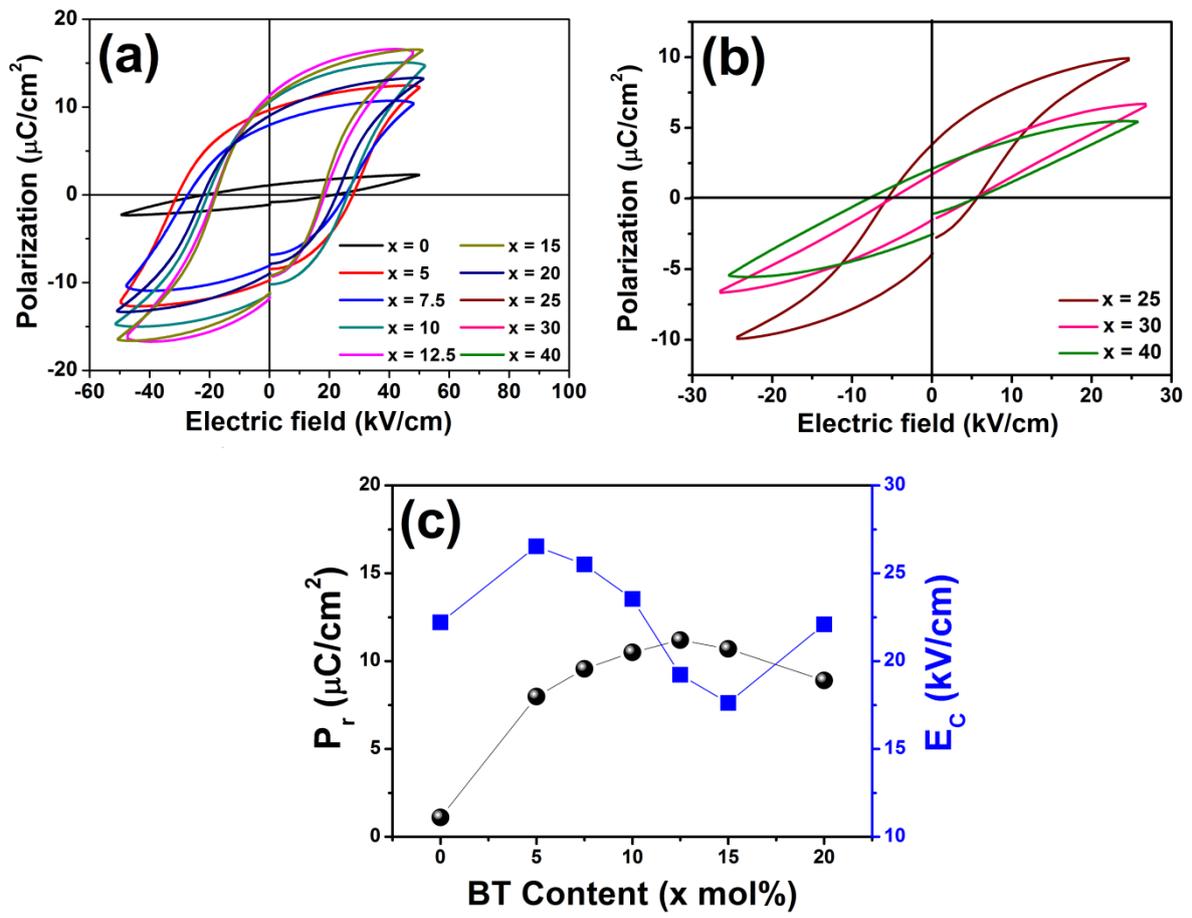

**Figure 7**

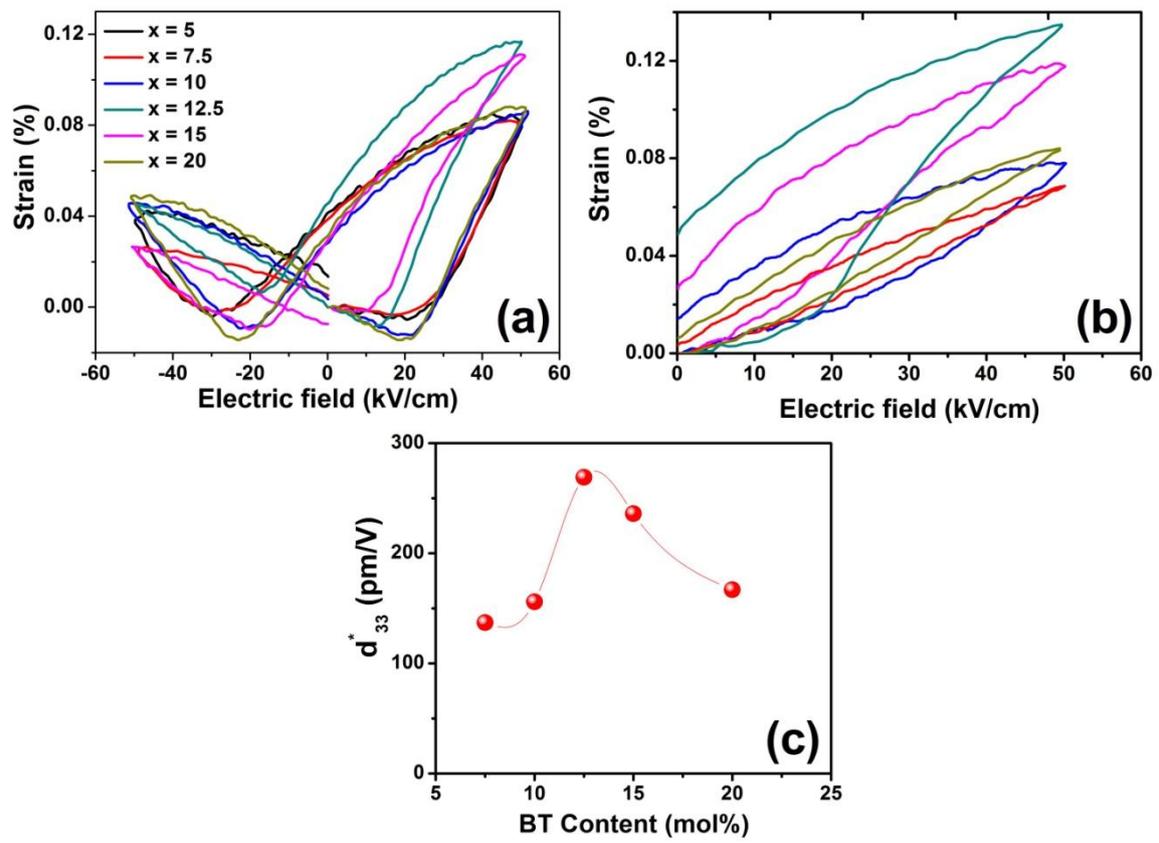

**Table 1**

| SG | *P m -3 m* | | *P 4 m m* | | | *P b m a* | | | |
|---|---|---|---|---|---|---|---|---|---|
| *x* | *a* (Å) | *V* (Å³) | *a=b*(Å) | *c* (Å) | *V* (Å³) | *a* (Å) | *b* (Å) | *c* (Å) | *V* (Å³) |
| 40 | 3.9799 | 63.0387 | | | | | | | |
| 30 | 3.9672 | 62.4389 | | | | | | | |
| 25 | 3.9596 | 62.0821 | | | | | | | |
| 20 | 3.9457 | 61.4265 | 3.9415 | 3.9332 | 61.1043 | | | | |
| 15 | | | 3.9346 | 3.9299 | 60.8326 | | | | |
| 12.5 | | | 3.9228 | 3.9443 | 60.6962 | | | | |
| 10 | | | 3.9257 | 3.9199 | 60.4090 | 5.5842 | 15.6365 | 5.5374 | 60.4390 |
| 7.5 | | | | | | 5.5760 | 15.6100 | 5.5290 | 60.1574 |
| 5 | | | | | | 5.5715 | 15.5895 | 5.5224 | 59.9580 |
| 0 | | | | | | 5.5566 | 15.5022 | 5.4973 | 59.1921 |

**Table 2**

| $x$ | Optimum sintering temperature ($^{o}$C) | Maximum relative density (%) |
|---|---|---|
| 0 | 1250 | 95.30±0.53 |
| 5 | 1185 | 96.69±0.46 |
| 7.5 | 1165 | 97.24±0.47 |
| 10 | 1150 | 97.55±0.47 |
| 12.5 | 1140 | 97.47±0.47 |
| 15 | 1120 | 97.19±0.47 |
| 20 | 1100 | 96.98±0.46 |
| 25 | 1100 | 96.81±0.42 |
| 30 | 1100 | 96.64±0.42 |
| 40 | 1120 | 96.49±0.42 |

**Table 3**

| $x$ (mol%) | $\varepsilon_r$ ($T_m$) | $\varepsilon_r$ (RT) | $\tan\delta$ (RT) | $P_r$ ($\mu C/cm^2$) | $E_C$ (kV/cm) | $d^*_{33} = S_{max}/E_{max}$ | $d_{33}$ (pC/N) | $k_p$ | $Q_m$ |
|---|---|---|---|---|---|---|---|---|---|
| 0    | 2703  | 174  | 0.026 | 1.10  | 22.2 | --- | 28 | 14 | 440 |
| 5    | 8433  | 304  | 0.065 | 7.98  | 26.5 | --- | 28 | 18 | 207 |
| 7.5  | 8929  | 303  | 0.131 | 9.56  | 25.5 | 137 | 33 | 11 | 267 |
| 10   | 10489 | 445  | 0.063 | 10.50 | 23.5 | 156 | 32 | 9  | 251 |
| 12.5 | 8842  | 795  | 0.081 | 11.20 | 19.2 | 269 | 30 | 12 | 162 |
| 15   | 6000  | 707  | 0.054 | 10.70 | 17.6 | 236 | 41 | 13 | 129 |
| 20   | 4678  | 978  | 0.076 | 8.90  | 22.1 | 167 | 49 | 10 | 228 |
| 25   | 2932  | 2400 | 0.033 | 3.70  | 5.5  | --- | --- | --- | --- |
| 30   | 2143  | 1874 | 0.037 | 1.60  | 4.9  | --- | --- | --- | --- |
| 40   | 1844  | 1190 | 0.031 | 2.00  | 4.9  | --- | --- | --- | --- |